\documentclass[twocolumn,english,showpacs,floatfix,amsmath,amssymb,prb,superscriptaddress]{revtex4-1}
\date{\today}

\usepackage{color} 

\usepackage{graphicx}
\usepackage{amsmath}
\usepackage{amssymb}

\newcommand{\be}{\begin{equation}}
\newcommand{\ee}{  \end{equation}}
\newcommand{\ba}{\begin{eqnarray}}
\newcommand{\ea}{  \end{eqnarray}}

\begin{document}

\title{Unveiling a crystalline topological insulator in a Weyl semimetal with time-reversal symmetry}

 \author{Liliana Arrachea} 
\affiliation{Departamento de F\'{\i}sica and IFIBA, Facultad de Ciencias Exactas y Naturales, Universidad de Buenos Aires, Pab.\ I, Ciudad Universitaria, 1428 Buenos Aires, Argentina }

 \author{Armando A. Aligia}
\affiliation{Centro At\'{o}mico Bariloche and Instituto Balseiro, Comisi\'{o}n Nacional
de Energ\'{\i}a At\'{o}mica, 8400 Bariloche, Argentina}

\date{\today}
\begin{abstract}
We consider a natural 
generalization of the lattice model for 
a periodic array of  two layers, $A$ and $B$, of 
spinless electrons proposed by Fu [Phys. Rev. Lett. {\bf 106}, 106802 (2011)]
as a prototype for a crystalline 
insulator. This model has time-reversal symmetry and broken inversion symmetry. 
We show that when the
intralayer next-nearest-neighbor hoppings $t_2 ^{a}, \; a= A,B$
vanish, this model supports a Weyl semimetal phase for a wide range of the remaining model parameters. 
When 
the effect of 
$t_2 ^{a}$
is considered, topological 
crystalline insulating phases take place within the Weyl semimetal one. 
By mapping to an effective Weyl Hamiltonian  we derive some  analytical results  
for the phase diagram as well as for  the structure of the nodes in the spectrum of
the Weyl semimetal.
\end{abstract}
\maketitle
\section{Introduction}
In recent years, it has become clear that 
topology plays a crucial role in classifying the phases of matter. The prelude of this important conceptual development took place in the '80s with the discovery of the quantum Hall effect in two-dimensional electron gases in strong magnetic fields \cite{1}.  
The fact that a pure magnetic field is not crucial to get a topological state in an electron lattice was later proposed by Haldane. \cite{haldane}
The recent advances were triggered by the theoretical proposal and subsequent experimental observation of two-dimensional (2D) topological insulators. These were originally regarded as extensions of the quantum Hall effect to time-reversal invariant systems which are subject to a strong spin-orbit interaction \cite{2,3,4}.
 Since then, a number of additional topological systems were proposed, including three-dimensional (3D) topological insulators \cite{fu-kane-mele,hasan}, topological superconductors, \cite{5,6,7,8,9,10,11,12,13} topological crystalline insulators (TCI), \cite{fu1,fu2} as well as Weyl semimetals (WSM). \cite{16,fang,17, ojanen,okugawa1,Murakami} 
 
 The role of crystal point symmetries, 
 usually  present in real solids, 
 to characterize the topological properties of the band structure was stressed in Ref. \onlinecite{fu1}, 
 where  the concept of 
 \textquotedblleft topological crystalline  insulator\textquotedblright  was introduced.  
 These are 3D systems which 
 have fourfold ($C_4$) or six-fold ($C_6$) rotational symmetries and display topologically non-trivial insulating phases with surface protected metallic states in high-symmetry directions. 
 Unlike other topological insulators,   spin-orbit coupling is not a crucial ingredient to drive the TCI phase. In fact, this phase could take place even in a spinless system.
 A prototypical model supporting the TCI phase was formulated in Ref. \onlinecite{fu1}
on the basis of a tight-binding Hamiltonian for spinless electrons. The
 possible realization of the TCI phase  in the compound SnTe, as well as the compounds PbTe and PbSe under pressure   was discussed in Ref. \onlinecite{fu2}. Interestingly, the
 possibility of realizing 2D Dirac fermions and the ''parity anomaly'' in PbTe had been previously suggested in an early work. \cite{panom}
 
 The underlying point group symmetry was also identified as the main ingredient to stabilize other topological properties 
 like the type of dispersion relation around the nodes of Weyl 
 semimetals. \cite{fang} 
 Unlike the topological insulators, which have a gap in the spectrum, WSMs are characterized by gapless points (Weyl nodes) in the Brillouin zone. Close to the nodes,
 the effective Hamiltonian is that of a 3D Weyl fermion. \cite{fang,17} As stressed in Ref. \onlinecite{ojanen}, the band touching at the nodes in these 3D systems is possible when inversion symmetry or time reversal symmetry is broken. The WSM phase provides the scenario for several exotic phenomena, like the so called 
 chiral anomaly in the presence of electric and magnetic fields, \cite{chiranomaly} and the existence of topologically protected surfaces and Fermi arcs in slab configurations. In Ref. \onlinecite{ojanen} a model with time-reversal symmetry based on a tight-binding Hamiltonian, containing a spin-orbit term but broken inversion symmetry  was studied as an example of a lattice model for a Weyl 
 semimetal. More recently, another model containing a WSM phase with time-reversal symmetry was analyzed. \cite{okugawa1} This model is  the tight-binding Hamiltonian with spin-orbit interaction, proposed by Fu, Kane and Mele \cite{fu-kane-mele}  as a prototype for a topological insulator. In  Ref. \onlinecite{okugawa1} it was considered in a 3D diamond lattice with the additional ingredient of a staggered on-site potential to break the inversion symmetry and the WSM is shown to take place in the middle of two topological insulating phases. 
 
 In the present work, we show that a natural generalization of the model introduced in Ref. \onlinecite{fu1} 
 as a prototype of a TCI, has a rich phase diagram, including also a WSM phase. Actually, we show that the latter phase can be regarded as the mother phase of the TCI one. 
 This model has the symmetry that corresponds to the space group $P_{4mm}$ and time-reversal symmetry. 
 The point group is $C_{4v}$ and therefore inversion symmetry is lacking. The  spin-orbit interaction is absent and the model has time-reversal symmetry.

 The paper is organized as follows. We present the model in Section II. It consists in a tight-binding Hamiltonian for spinless electrons with two layers and two orbitals per unit cell, very similar to the one proposed by Fu in Ref. \onlinecite{fu1}. The features of the band structure in the different expected phases are summarized in Section III. In Section IV we consider a limit of the model where the in-plane next-nearest neighbor hopping parameters vanish. In this limit, the effective low energy model
 is a generalized Weyl Hamiltonian, which can be exactly solved. We find the phase diagram in this limit and analyze the structure of the spectrum in the bulk as well as in a slab configuration. We show that within a wide range of parameters there exist a WSM phase with nodes in 3D and Fermi arcs in the slab, as well as a normal insulating phase. In Section  V we analyze the role of the next-nearest neighbor hopping parameter and find that this ingredient drives TCI phases within the WSM one. Finally, we present a summary in Section VI and we discuss which materials  are candidates to realize the present model and phases.
 
 \begin{figure}[!h]
\begin{center}
\includegraphics[width=0.45\textwidth]{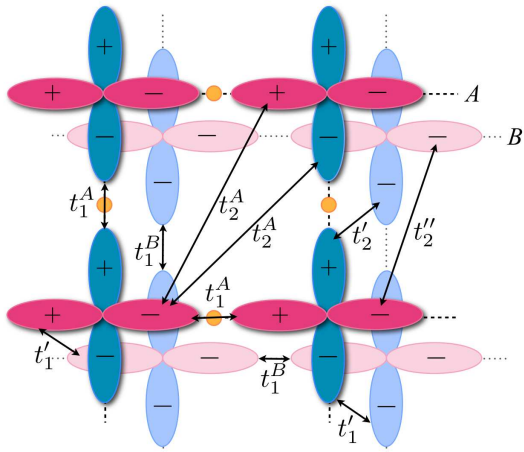}
\caption{(Color online) Sketch of a bilayer with planes $A$ and $B$ indicating the intra-plane and inter-plane hopping elements considered in the model.
 The sign convention for the $d_{xz}$ and $d_{yz}$ (or  $p_x$ and $p_y$) orbitals is also indicated. The circles in the plane $A$ indicate the possible presence 
 of intermediate atoms with orbitals that hybridize with the neighboring ones, renormalizing the hopping elements. }
\label{fig1}
\end{center} 
\end{figure}
 
 \section{Model} \label{Model}
 We consider a tight-binding model associated to two orbitals $d_{xz}$ and $d_{yz}$ or $p_x$ and $p_y$  in a tetragonal lattice with two atoms as in 
 Ref. \onlinecite{fu1}.  
 The ensuing Hamiltonian is
 \be
 H=\sum_l \left( H^A_l + H^B_l + H^{AB}_l \right).
 \ee
 where $l$ labels  a couple of layers, $A$ and $B$ along the $z$ axis. The two layers are described by the Hamiltonians $H^A_l$ and $H^B_l$, respectively,
  while $H^{AB}_l$ describes the inter-layer hybridization terms. 
Explicitly, the terms in the ensuing effective Hamiltonian read
 \ba
 H^a_l &=& \sum_{i,j} \sum_{\alpha,\beta}c^{\dagger}_{a,\alpha}({\bf r}_i,l) t^a_{\alpha, \beta} ({\bf r}_i-{\bf r}_j)  c_{a,\beta}({\bf r}_j,l), \; a= A,B\nonumber \\
H^{AB}_l & = & \sum_{i,j} \sum_{\alpha, \beta} t^{\prime}_{\alpha, \beta} ({\bf r}_i-{\bf r}_j) \left[ c^{\dagger}_{A, \alpha} ({\bf r}_ i, l)  c_{B, \beta} ({\bf r}_ j, l) + H. c. \right] \nonumber \\
& & + t^{\prime}_z \sum_i \sum_{\alpha} \left[ c^{\dagger}_{A, \alpha} ({\bf r}_i, l)  c^{\dagger}_{B, \alpha} ({\bf r}_i, l+1) + H. c. \right].
\ea
The parameter $t^a_{\alpha, \beta}({\bf r}_i-{\bf r}_j), \; a= A,B$ denotes the intra-plane hopping matrix element between the orbitals $\alpha$ and $\beta$ localized at the
atomic positions ${\bf r}_i$ and ${\bf r}_j$, with ${\bf r}=(x,y)$, on the plane. Keeping hopping elements up to next-nearest neighbors $\left( i,j \right)$ leads to the following 
Fourier transform of $H$,
\[ H({\bf k}) = \left( \begin{array}{cc}
H^A({\bf k}) & H^{AB}({\bf k})   \\
 H^{AB}({\bf k})^{\dagger} &H^B({\bf k})   \end{array} \right),\] 
with 
\[\begin{array}{cc} 
& H^a({\bf k}) =    2 t_1^{a} \left( \begin{array}{cc}
\cos k_x & 0   \\
 0  &\cos k_y  \end{array} \right) +  \\
 &  2 t_2^a \left( \begin{array}{cc}
\cos k_x \cos k_y & \sin k_x \sin k_y  \\
 \sin k_x \sin k_y   &\cos k_x \cos k_y  \end{array} \right), \end{array}
 \] 
and
\[\begin{array}{cc} 
& H^{AB}({\bf k}) =   ( t^{\prime}_1 + t^{\prime}_z e^{i k_z} )\left( \begin{array}{cc}
1 & 0   \\
 0  & 1 \end{array} \right) +  \\
 &  2 \left( \begin{array}{cc}
t^{\prime}_2 \cos k_x + t^{\prime \prime}_2  \cos k_y & 0  \\
 0 &t^{\prime}_2 \cos k_y + t^{\prime \prime}_2  \cos k_x   \end{array} \right). \end{array}
 \] 
The different hopping processes are indicated in Fig. \ref{fig1}. 
The minimal model for a TCI considered in Ref. \onlinecite{fu1} corresponds to 
$t^{\prime}_2=t^{\prime \prime}_2$. Notice that 
 because of symmetry, the hopping elements along the $x$ and $y$ directions are non-vanishing 
only for hopping processes between the same type of orbitals. 
For example,  $d_{xz}$ and $d_{yz}$ (or $p_x$ and $p_y$) orbitals at sites with the same $y$ coordinates, have opposite parities under 
a reflection through the $xz$ plane.
The hopping elements
$t^{\prime}_2$ and $t^{\prime \prime}_2$ correspond to the hopping between $d_{\alpha z}$ (or $p_{\alpha}$) orbitals of different $z$  planes aligned in the 
$\alpha$ direction or in the perpendicular one respectively.
In the general case, we expect
$t^{\prime}_2 > t^{\prime \prime}_2$. 

We show that for such case and other parameters close or identical to those considered 
in Ref. \onlinecite{fu1}, the present model  exhibits a WSM phase with Fermi arcs
and Weyl nodes. A crucial ingredient is  that the signs 
of $t_1^A$ and  $t_1^B$ are opposite. 
This could happen if
for example in the A plane, the effective hopping between $d_{\alpha z}$ orbitals is originated 
by a second-order process through
intermediate occupied $p_\alpha$ orbitals of an atom lying in the middle, as shown in Fig. 1 
(a very usual case in perovskites of transition metals and oxygen),
while in the B plane, the intermediate occupied orbitals are $s$ 
instead of the $p_\alpha$ ones. Further discussion on this point is deferred to Section \ref{sum}.

\section{Phases and spectral properties}
The band structure   of $H({\bf k})$ indicates the existence of three possible phases in this model, depending on  the ratio between the different hopping elements. 

\subsection{Normal insulator}
 \begin{figure}[!h]
\begin{center}
\includegraphics[width=0.45\textwidth]{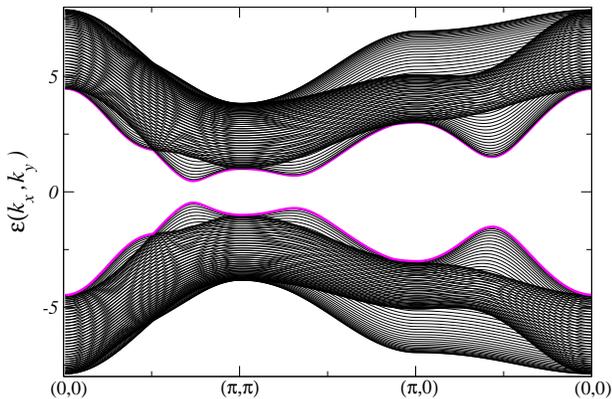}
\caption{(Color online) Band structure of the slab with $N=40$ bilayers in the NI phase. Parameters are $t_1^A=1=-t_1^B=1$, $t_2^A=0.5= -t_2^B$, $t_1^{\prime}=3.5$, $t_z^{\prime}=2$ 
and $t_2^{\prime}=0.8$, $t_2^{\prime \prime}=0.1$ States of the surfaces are shown in thick magenta lines. States of the surfaces with energies at the boundary of the gap are shown in thick magenta lines.  }
\label{fig2a}
\end{center} 
\end{figure}

The normal insulator (NI) phase is characterized by a spectrum with a gap in all the ${\bf k}=(k_x,k_y,k_z)$ points of the first 3D Brillouin zone as well as a spectrum with gap for a slab configuration with 
$N$ bilayers. An example is shown in Fig. \ref{fig2a}.

\subsection{Topological crystalline insulator}
 \begin{figure}[!h]
\begin{center}
\includegraphics[width=0.45\textwidth]{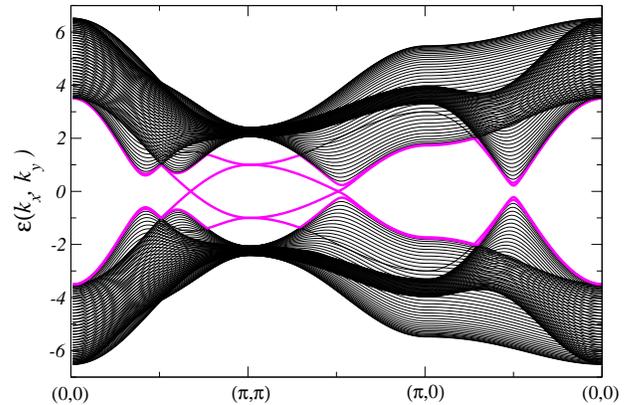}
\caption{ (Color online) Band structure of the slab  $N=40$ bilayers in the TCI phase. 
Parameters are $t_1^A=1=-t_1^B=1$, $t_2^A=0.5= -t_2^B$, $t_1^{\prime}=2.$, $t_z^{\prime}=2$ 
and $t_2^{\prime}=0.8$, $t_2^{\prime \prime}=0.1$. States of the surfaces with energies within or at the boundary of the bulk gap are shown in thick magenta lines. }
\label{fig2b}
\end{center} 
\end{figure}
 The topological crystalline insulator (TCI) phase is the one introduced in Ref. \onlinecite{fu1}. It is also characterized by a gapped spectrum in the bulk but metallic surface states along the direction $(001)$, which preserves $C_4$ symmetry.
These states are doubly degenerate at the high-symmetry point $\overline{M}= (\pi,\pi)$ of the square 2D projected  Brillouin zone. An example is shown in Fig. \ref{fig2b}.

 \subsection{Weyl semimetal} 
\begin{figure}[!h]
\begin{center}
\includegraphics[width=0.45\textwidth]{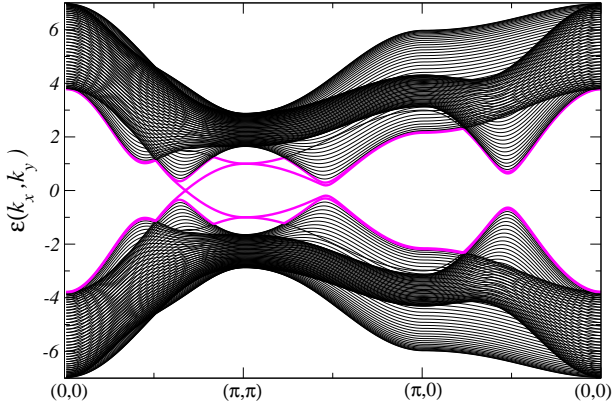}
\caption{(Color online) Band structure of the slab  $N=40$ bilayers in the WSM phase. Parameters are $t_1^A=1=-t_1^B=1$, $t_2^A=0.5= -t_2^B$, $t_1^{\prime}=2.5$, $t_z^{\prime}=2$ 
and $t_2^{\prime}=0.8$, $t_2^{\prime \prime}=0.1$. States of the surfaces with energies within or at the boundary of the bulk gap are shown in thick magenta lines. }
\label{fig2c}
\end{center} 
\end{figure}

 The Weyl semimetal (WSM) phase is characterized by  the existence of nodes in the bulk spectrum, where two bands touch each other. As stressed in Ref. \onlinecite{ojanen} this phase 
 exists in systems with broken time-reversal symmetry or broken spacial inversion symmetry. The present model is formulated for spinless electrons, thus time-reversal symmetry is trivially preserved. However, spacial inversion symmetry is
 broken along the $z$ axis. 
 In the present model the nodes are points  located at the plane $k_z=0$ and $k_z=\pi$ of the 3D Brillouin 
 zone. These nodes appear in pairs
  and define vertices of Dirac cones, in which neighborhood the dispersion relation of the touching bands is in general linear. For nodes lying on 
 symmetry points the dispersion can be also quadratic, as discussed in the next section. This phase also support metallic surfaces in the slab configuration. 
 The projected Fermi surface of these metallic state defines Fermi arcs  on the 2D square Brillouin zone, which connect the nodes of the bulk spectrum.
 This feature will be further discussed in Section IV.
 An example of the band structure in the slab is shown in Fig. \ref{fig2c}. 
 Metallic surface states are clearly distinguished and the existence of Fermi arcs are inferred by noticing 
 that the closed triangular path  
 along the 2D projected Brillouin zone 
 $ {\cal P}: \{(0,0) \rightarrow (\pi,\pi) \rightarrow (\pi,0) \rightarrow (0,0)\}$, chosen to draw the Fig. \ref{fig2c}, the number of states of the Fermi surface is $2 n_F$, with $n_F$ odd for a Fermi energy close to zero ($n_F=1$ in the example of the Fig. 2). Hence, the Fermi surface corresponding to the states of a given slab surface defines an open arc
 connecting two Dirac cones of the bulk spectrum. 
 The two slab surfaces define arcs with different concavities. Thus, the joint states of the pair of slab surfaces define a closed Fermi surface.    
 Instead, within the TCI phase, $n_F$ is even for the metallic surface states (see for instance Fig. \ref{fig2b} where $n_F=2$). This is due to the fact that in the TCI phase, each slab surface 
 has states forming a closed Fermi surface.

 \section{Effective Weyl Hamiltonian for $t_{2}^A=t_2^B=0$.}
 In order to identify the ingredients that define the nature of the spectrum in the different phases, let us focus on the case with $t_1^A=-t_1^B=t_1$ and 
 in the case where the terms proportional to
 $t_2^A, t_2^B $ vanish. The latter condition is satisfied for any $t_2^A$ and $t_2^B$ if we focus on ${\bf k}$-points with
  $(k_x,k_y)=(\pm\pi/2,\pi), (\pi, \pm \pi/2), (0, \pm \pi/2), (\pm \pi/2,0)$. Without such terms $H({\bf k})$ is blocked in two matrices,
 $H_{x}({\bf k})$ and $H_{y}({\bf k})$ corresponding to the subspaces related to the $d_{xz}$ and $d_{yz}$ (or $p_x$ and $p_y$) orbitals, respectively. 
 Hereafter, we will simply label these orbitals with the index $\alpha=x, y$, respectively. They read
 \be \label{weyl}
 H_{{\alpha}}({\bf k})= g_x^{\alpha} \sigma_x+ g_y^{\alpha} \sigma_y+g_z^{\alpha} \sigma_z,
 \ee
 with $\alpha=x,y$ where $\sigma_{x,y,z}$ are Pauli matrices, while
 \ba \label{ges}
 g_x^{\alpha}({\bf k}) & = & t_1^{\prime}+ t_z^{\prime} \cos k_z + 2 \left( t^{\prime}_2 \cos k_{\alpha}+ t^{\prime \prime}_2 \cos k_{\overline{\alpha}} \right) \nonumber \\
 g_y^{\alpha}({\bf k}) & = & -t^{\prime}_z \sin k_z,\nonumber \\
  g_z^{\alpha}({\bf k}) & = & 2 t_1 \cos k_{\alpha}, 
  \ea
 where we have introduced the notation $\overline{x}= y$ and $\overline{y}= x$.  The Hamiltonian (\ref{weyl}) would coincide with Weyl Hamiltonian
 if $g_{j}^{\alpha}= k_{j}$, with $j=x,y,z$. 
 The eigenenergies of the Hamiltonians $H_{{\alpha}}({\bf k})$ are
 \be
 \varepsilon_{\pm}^{\alpha}({\bf k}) =\pm \sqrt{ \left(g_x^{\alpha}\right)^2+\left(g_y^{\alpha}\right)^2+\left(g_z^{\alpha} \right)^2}.
 \ee
 
\subsection{Nodes in the spectrum. Weyl semimetal}
 The Weyl semimetal phase takes place when the spectrum is gapless for values ${\bf K}$ satisfying simultaneously $g_j({\bf K})=0$.  The ${\bf K}$ points where the bands touch receive the name of nodes.
 We can identify  two different kind of nodes, which are described below.

\subsubsection{Nodes on symmetry  points}
We can easily identify four examples of such points. The first one is ${\bf K}_1=(\pm \pi/2, \pi, \pi)$.
In that case, expressing ${\bf k}= {\bf K}_1+ {\bf q}$ and performing a Taylor expansion for small ${\bf q}$, we find 
\ba \label{g1}
g_x^x({\bf K}_1+ {\bf q})&=& \Delta_1- 2 t^{\prime}_2 q_x + t_2^{\prime \prime} q_y^2,\nonumber \\
g_y^x({\bf K}_1+ {\bf q})  & = & t_z^{\prime} q_z, \nonumber \\
g_z^x({\bf K}_1+ {\bf q}) & = & -2 t_1 q_x,
\ea
with $ \Delta_1= t_1^{\prime}-t^{\prime}_z - 2 t_2^{\prime \prime}$.
Hence $\varepsilon_{\pm}^{x} ({\bf K}_1+ {\bf q})$
is gapless for a combination of the hopping parameters $t_1^{\prime}, t^{\prime}_z, t_2^{\prime \prime}$ satisfying $\Delta_1=0$, instead
$\varepsilon_{\pm}^{y}({\bf K}_1+ {\bf q}))$ are gapped.  The opposite situation takes 
place for ${\bf K}_1^{\prime}= (\pi, \pm \pi/2,\pi)$, 
in which case
\ba \label{g1p}
g_x^y({\bf K}_1^{\prime}+ {\bf q}) &=& \Delta_1 - 2 t^{\prime}_2 q_y + t_2^{\prime \prime} q_x^2,\nonumber \\
g_y^y({\bf K}_1^{\prime}+ {\bf q}) & = & t_z^{\prime} q_z, \nonumber \\
g_z^y({\bf K}_1^{\prime}+ {\bf q}) & = & -2 t_1 q_y,
\ea
Hence $\varepsilon_{\pm}^{y} ({\bf K}_1^{\prime}+ {\bf q})$
is gapless for the  combination of the hopping parameters $t_1^{\prime}, t^{\prime}_z, t_2^{\prime \prime}$ satisfying $\Delta_1=0$, while 
$\varepsilon_{\pm}^{x} ({\bf K}_1^{\prime}+ {\bf q})$ are gapped.

The second example corresponds to points close to ${\bf K}_2=(\pm \pi/2,0,\pi)$. Performing a Taylor expansion of the coefficients $g$ around 
these points we get
\ba \label{g2}
g_x^x({\bf K}_2+ {\bf q})&=& \Delta_2 - 2 t^{\prime}_2 q_x - t_2^{\prime \prime} q_y^2,\nonumber \\
g_y^x({\bf K}_2+ {\bf q}) & = & t_z^{\prime} q_z, \nonumber \\
g_z^x({\bf K}_2+ {\bf q}) & = & -2 t_1 q_x,
\ea
with $ \Delta_2= t_1^{\prime}-t^{\prime}_z + 2 t_2^{\prime \prime}$.
Hence $\varepsilon_{\pm}^{x} ({\bf K}_2+ {\bf q})$
is gapless for a combination of the hopping parameters $t_1^{\prime}, t^{\prime}_z, t_2^{\prime \prime}$ satisfying $\Delta_2=0$, while 
$\varepsilon_{\pm}^{y}({\bf K}_2+ {\bf q})$ are gapped.  Similarly,  the 90-degree rotated point ${\bf K}_2^{\prime}= (0, \pm \pi/2,\pi)$ 
corresponds to a gapless point for the $y$-bands when $\Delta_2=0$.

The third case corresponds to ${\bf K}_3=(\pm \pi/2, \pi, 0)$. Close to this point,  the Taylor expansion cast
\ba \label{g3}
g_x^x({\bf K}_3+ {\bf q}) & = & \Delta_3  - 2 t^{\prime}_2 q_x + t_2^{\prime \prime} q_y^2, \nonumber \\
g_y^x({\bf K}_3+ {\bf q}) & = & - t_z^{\prime} q_z, \nonumber \\
g_z^x({\bf K}_3+ {\bf q}) & = & -2 t_1 q_x,
\ea
with $ \Delta_3 = t^{\prime}_1 + t^{\prime}_z - 2 t^{\prime \prime}_2$, which leads to gapless states in $\varepsilon_{\pm}^{x} ({\bf K}_3+ {\bf q})$ when $ \Delta_3 = 0$, 
and gapped $\varepsilon_{\pm}^{y} ({\bf K}_3+ {\bf q})$ bands. Instead, the latter are gapless for ${\bf K}_3^{\prime}=(\pi , \pm \pi/2 , 0)$, when $ \Delta_3 = 0$.

Similarly, a  Taylor expansion  around ${\bf K}_4=(\pm \pi/2, 0, 0)$ cast
\ba \label{g4}
g_x^x({\bf K}_4+ {\bf q}) & = & \Delta_4 - 2 t^{\prime}_2 q_x - t_2^{\prime \prime} q_y^2, \nonumber \\
g_y^x({\bf K}_4+ {\bf q}) & = & - t_z^{\prime} q_z, \nonumber \\
g_z^x({\bf K}_4+ {\bf q}) & = & -2 t_1 q_x,
\ea
with $ \Delta_4 = t^{\prime}_1 + t^{\prime}_z + 2 t^{\prime \prime}_2$, implying
gapless states in $\varepsilon_{\pm}^{x} ({\bf K}_4+ {\bf q})$ when $ \Delta_4 = 0$, 
and gapped $\varepsilon_{\pm}^{y} ({\bf K}_4+ {\bf q})$ bands. The latter bands are gapless at  ${\bf K}_4^{\prime}=(0 , \pm \pi/2 , 0)$,
for $\Delta_4=0$.

The different gapless points define nodes in the spectrum and are indicated in Fig. \ref{figmap}. For parameters leading to $\Delta_j=0, \; j=1, \ldots, 4$, the system is in a 
Weyl semimetal phase with metallic surfaces in a slab geometry and Fermi arcs. Notice, however, that
the effective Hamiltonian is not the conventional Weyl Hamiltonian. Furthermore, the effective dispersion relations $\varepsilon_{\pm}^{x} ({\bf K}_j+ {\bf q})$ are linear in $q_z$ and $q_x$ while they are quadratic in  $q_y$. Instead for $\varepsilon_{\pm}^{y} ({\bf K}_j^{\prime}+ {\bf q})$ 
 they are linear in $q_z$ and $q_y$ while they are quadratic in  $q_x$. The latter peculiarity is due to the fact that 
${\bf K}_j, \; j=1, \ldots, 4$ are invariant under a reflection $k_y \rightarrow -k_y$ while ${\bf K}_j^{\prime},\; j=1, \ldots, 4$,  are invariant under the operation
$k_x \rightarrow -k_x$. In Ref. \onlinecite{fang} the possibility of multiple Weyl points was proposed to take place in nodes located at high symmetry points of the reciprocal lattice, 
which remain invariant under operations of the $C_n$ point group. In the
present case we find that the inversion symmetry may cause a more exotic scenario, 
with an anisotropic dispersion relation changing from quadratic to linear depending on the direction.

\subsubsection{Nodes away from high-symmetry points}
For $t_{2}^{A}=t_{2}^{B}=0$, and for sets of the remaining hopping parameters that do not satisfy
$\Delta_j=0$, the spectrum can still have nodes as long as 
$\Delta_j/t^{\prime \prime}_2 <0,\; j=1,3$ or $\Delta_j/t^{\prime \prime}_2 >0,\; j=2,4$. Such nodes can be actually regarded as splittings of the
nodes placed at the points $\mathbf{K}_{j}$ and 
$\mathbf{K}_{j}^{\prime }$ described in the previous section into pairs that displace along the symmetry lines. Hence, the number of nodes in this case
is twice the number of nodes for $\Delta_j=0$. 

To give a specific example, when $\Delta_1 \neq 0$, with $\Delta_1/t^{\prime \prime}_2 <0$, the node at $\mathbf{K}_{1}$ splits  in two new nodes
at the positions  $\mathbf{K}_{1,\pm }=(\pi
/2,\pm k_{1},\pi )$, with $k_{1}=$ $|\arccos \left[ 
|\Delta_{1}/(2t_{2}^{\prime \prime })|-1\right] |.$
The new $g$-parameters of the Weyl Hamiltonian are 
\begin{eqnarray}
g_{x}^{x}(\mathbf{K}_{1,\pm }+\mathbf{q}) &=&-2t_{2}^{\prime }\left[ q_{x}\mp \sin
(k_{1})q_{y} \right],  \notag \\
g_{y}^{x}(\mathbf{K}_{1,\pm }+\mathbf{q}) &=&t_{z}^{\prime }q_{z},  \notag \\
g_{z}^{x}(\mathbf{K}_{1,\pm }+\mathbf{q}) &=&-2t_{1}q_{x}.
\end{eqnarray}
Interestingly, the effective Hamiltonian at the new nodes is linear in $\mathbf{q}$. This is a consequence of the fact that  these nodes lie at points
 which do not have any particular symmetry.

\begin{figure}[!h]
\begin{center}
\includegraphics[width=8.cm]{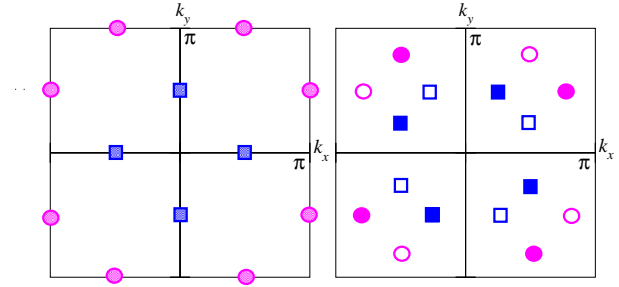}
\caption{(Color online) Nodes in the Weyl semimetal phase in the $(k_x,k_y)$ plane. Left: Nodes for $\Delta_j=0$. Circles correspond to $\Delta_1=t_1^{\prime}-t^{\prime}_z - 2 t_2^{\prime \prime} =0$ for $k_z= \pi$ or
$\Delta_3=t^{\prime}_1 + t^{\prime}_z - 2 t^{\prime \prime}_2=0$ for $k_z= 0$ (Points with coordinates $k_x= \pm \pi$, $k_y=\pm \pi$ are equivalent but they are indicated for clarity). 
Squares correspond  to 
$\Delta_2=t_1^{\prime}-t^{\prime}_z + 
2 t_2^{\prime \prime} =0$ for $k_z= \pi$, or $ \Delta_4 = t^{\prime}_1 + t^{\prime}_z + 2 t^{\prime \prime}_2$ for $k_z= \pi$
or $ \Delta_4 = t^{\prime}_1 + t^{\prime}_z + 2 t^{\prime \prime}_2=0$ for $k_z= 0$. Dark (light) symbols correspond to the touching of
the bands with $x$ ($y$) character.  Right: Nodes for $\Delta_j \neq 0$. Each of the nodes of the left panel splits into a pair of new nodes associated to monopoles with positive (negative) charges, corresponding to dark (white) symbols, respectively. }
\label{figmap}
\end{center} 
\end{figure}

For the remaining points, a similar analysis can be made. A sketch of the map of pairs of nodes is shown in Fig. \ref{figmap}.

\subsection{Monopoles of the Weyl semimetal}

One of the most interesting features associated to the Weyl Hamiltonian is
the underlying structure of monopoles associated to the nodes. This
structure has remarkable consequences in the electromagnetic response of
these systems, as described in Ref. \onlinecite{chiranomaly}. The emergence of
monopoles is associated to the Berry curvature. The latter is the vector field 
\begin{equation}
\mathbf{\Omega }(\mathbf{q})=\nabla _{\mathbf{q}}\times \mathbf{A}(\mathbf{q}),
\end{equation}
where 
\begin{equation}
\mathbf{A}(\mathbf{q})=i\langle \Psi _{-}(\mathbf{q})|\nabla _{\mathbf{q}}|\Psi _{-}(\mathbf{q})\rangle 
\end{equation}
is the Berry connection. \cite{book1,book2} The ket $|\Psi _{-}(\mathbf{q})\rangle $ 
corresponds to the ground state of the Weyl Hamiltonian. For a
linear relation between $\mathbf{g}$ and $\mathbf{q}$, the field $\mathbf{\Omega }$ 
corresponds to a monopole at the origin $\mathbf{q}=0$,
corresponding to a charge density $\rho (\mathbf{q}))=\mbox{sg}\{J\}\delta
^{3}(\mathbf{q})$, with \cite{Murakami,young,okugawa1} 
\begin{equation}
J=\mbox{Det}\left[ \frac{\partial (g_{x}^{x},g_{y}^{x},g_{z}^{x})}{\partial
(q_{x},q_{y},q_{z})}\right] .
\end{equation}%
It can be verified by explicitly computing $J$, that the charges of each of
the pairs $\mathbf{K}_{j,\pm }$ and $\mathbf{K}_{j,\pm }^{\prime }$ are
opposite. 
In fact, it is easy to see that $J \propto \mp \sin(k_{j})$. 
Hence, the limit $\Delta _{j}=0$ where the pairs merge into the
nodes $\mathbf{K}_{j}$ and $\mathbf{K}_{j}^{\prime }$ corresponds to the
annihilation of the  two opposite charges of the pairs of nodes.

 \subsection{Fermi arcs of the Weyl semimetal}
  \begin{figure}[!h]
\begin{center}
\includegraphics[width=9.5cm]{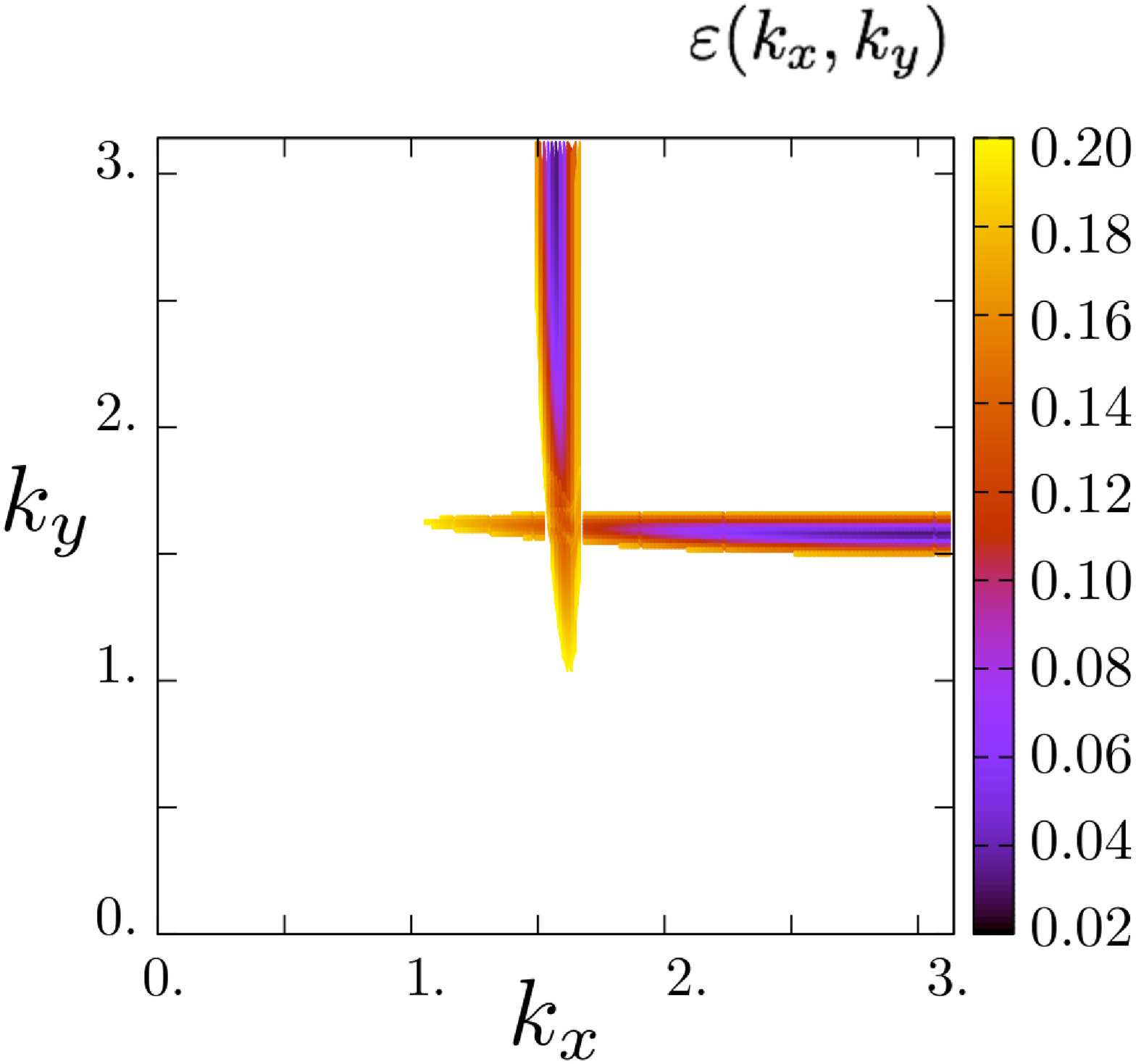}
\includegraphics[width=9.5cm]{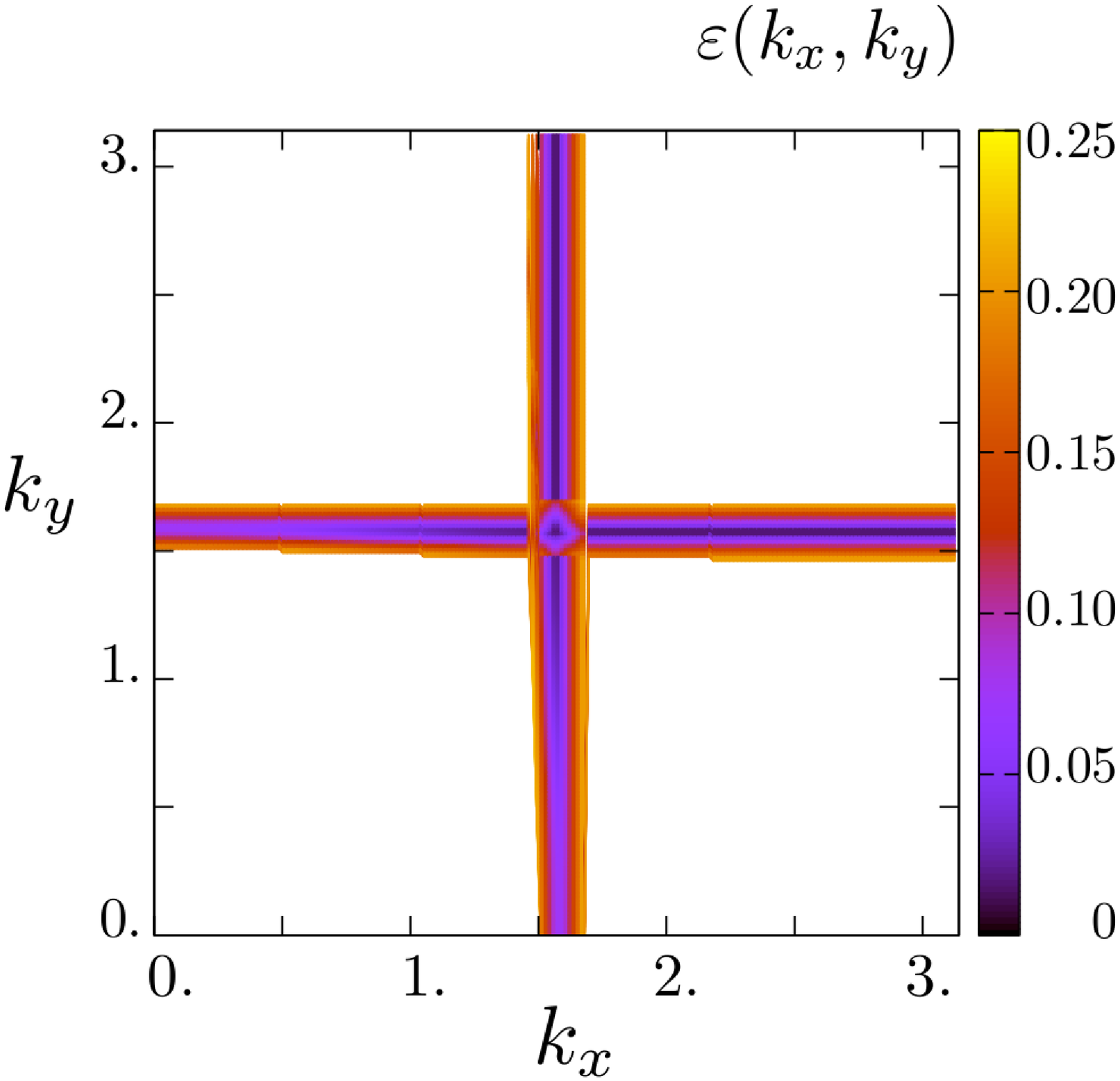}
\caption{(Color online) Map of the low energy sector of the band of the slab within the WSM phase 
with $t^{\prime \prime}_2>0$. 
Upper panel:  $t_1^A=1=-t_1^B=1$, $t_2^A= -t_2^B=0$, $t_1^{\prime}=2.5$, $t_z^{\prime}=2$ 
and $t_2^{\prime}=0.8$, $t_2^{\prime \prime}=0.1$, i.e. parameters satisfying
$0<t_z^{\prime}<t_1^{\prime} <t_z^{\prime}+2t^{\prime \prime}_2$. Lower panel: $t_1^{\prime}=1$ and other parameters like in 
the upper panel (notice that in this case
$0<t_1^{\prime} < t_z^{\prime}$).  
Only the first quadrant of the Brillouin zone is shown.}
\label{fermi}
\end{center} 
\end{figure}
Another remarkable characteristic of the Weyl semimetal phase is the existence of Fermi arcs when the model is considered in a slab configuration.
As stressed in previous works, \cite{ojanen,okugawa1} the arcs extend on the projected 2D Fermi surface, connecting nodes of the bulk spectrum.
In Fig. \ref{fermi} we illustrate the Fermi arcs  showing examples of map plots in the slab corresponding to Fermi energies close to zero for a value of $t^{\prime \prime}_2>0$.  
The upper panel corresponds to 
a particular set of parameters in the region $0<t_z^{\prime}<t_1^{\prime} <t_z^{\prime}+2t^{\prime \prime}_2$, 
for which $\Delta_1<0$. For these parameters, the nodes of the bulk are placed in positions like 
the ones indicated with circles in Fig. \ref{figmap}. 
One arc of the Fermi surface of the slab extends from one of these nodes, say $(\pi/2,k_1)$ to the node $(\pi/2,-k_1)$ 
(equivalent to the former by reflection symmetry) passing through $(\pi/2,\pi)$. See Fig. \ref{fermi}. 
The other arc shown in the figure is obtained by reflection through the line $k_x=k_y$.
For $t_z^{\prime}=t_1^{\prime}$ the node depicted with the dark circle evolving from the right of the first quadrant 
by increasing $t_z^{\prime}$
coincides with the white circle evolving from above at 
$(\pi/2,\pi/2)$, while for $t_z^{\prime}> t_1^{\prime}$ the circles of the Fig. \ref{figmap} 
move to $(k_x,k_y)$ coordinates similar to those of the square symbols of that figure. 
The Fermi surfaces, thus, cross at $ (\pi/2,\pi/2)$ as indicated in the lower panel of Fig. \ref{fermi}.

\subsection{Gapped spectrum. Normal insulator}
The analysis of Sections IV A to C reveals the existence of a Weyl semimetal (WSM) phase for a wide range of 
parameters $t^{\prime}_1, t^{\prime}_z, t^{\prime \prime}_2$, provided that
$t_1^{A}=-t_1^{B}$ and $t_{2}^{A}=t_{2}^B=0$, and irrespectively of the value of $t^{\prime}_2$. 
For a given $t^{\prime \prime}_2$, this phase extends in the $t_1^{\prime}, \; t_z^{\prime}$ plane 
from parameters consistent with $\Delta_j=0$, in which case the
nodes lye at symmetry points (see the left panel of Fig. \ref{figmap}),
along a wide region where these nodes split into pairs and displace along the different symmetry axis 
(like in the right panel of Fig. \ref{figmap}). Nodes from different pairs cross one another 
at the points  $(\pm \pi/2, \pm \pi/2)$ and continue their vertical or horizontal displacements,  
provided that the conditions $\Delta_j/t^{\prime \prime}_2 <0, j=1,3$ and $\Delta_j /t^{\prime \prime}_2>0, j=2,4$ 
are satisfied. For parameters  
that do not satisfy this condition, the spectrum is gapped for every ${\bf k}$ and the system becomes a 
normal insulator. The boundaries separating the WSM and NI phases are defined by the lines
$|t^{\prime}_1|= \left| 2 \left| t^{\prime \prime}_2 \right|+ t^{\prime}_z \right|$. The full phase diagram is shown in
 Fig. \ref{pd1}.

\section{Phase diagram for $t_{2}^A=-t_2^B\neq0$ }
 \begin{figure}[!h]
\begin{center}
\includegraphics[width=8.cm]{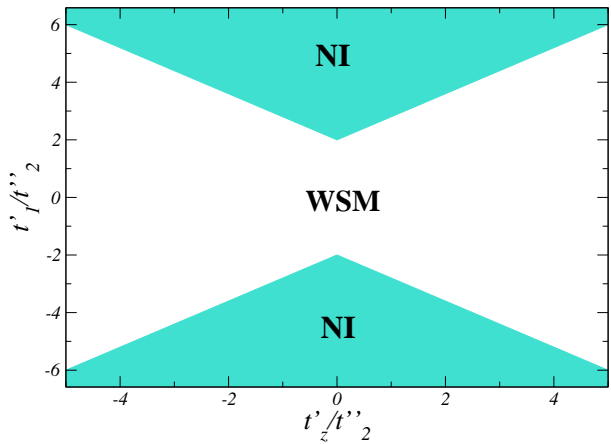}
\caption{Phase diagram for $t_{2}^A=-t_{2}^B=0$. NI and WSM denote, respectively, normal insulating phase and Weyl semimetal phase.}
\label{pd1}
\end{center} 
\end{figure}
\begin{figure}[!h]
\begin{center}
\includegraphics[width=8.cm]{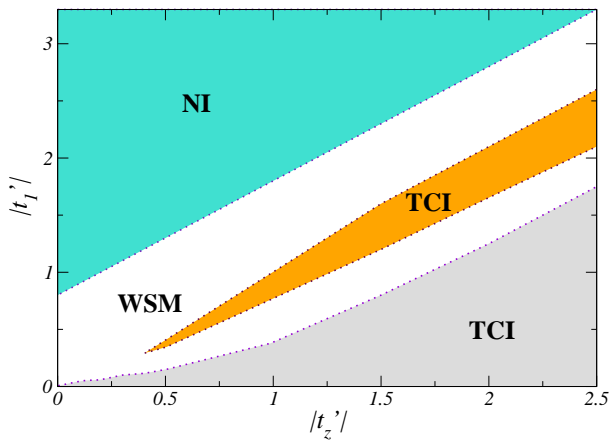}
\caption{Phase diagram for $t_{2}^A=-t_{2}^B=t_2=0.8$, $t_{1}^A=-t_{1}^B=1$,  $t^{\prime}_2=0.8$, $t^{\prime \prime}_2=0.1$. NI, TI and WSM denote, respectively, normal insulating phase, topological insulator and Weyl semimetal phase.}
\label{pd2}
\end{center} 
\end{figure}

The phase diagram for $t_{2}^A=t_2^B=0$ under the only condition $t_1^A=-t_1^B$ was analyzed in the previous section 
and it is shown in Fig. \ref{pd1}. 
The boundary between the WSM and NI phases could be calculated analytically in that case. 
We now turn to analyze the phase diagram for finite $t_2^A=- t_2^B =t_2 \neq 0$ on the basis of numerical calculations. 
In particular, we have solved the Hamiltonian in the bulk as well as in the slab and analyzed the structure of 
the spectrum. As mentioned in Section III the WSM phase is identified by nodes in the bulk spectrum. In addition, the number 
$2 n_F$ of intersections of the Fermi surface with energy close to zero with the path $ {\cal P}: \{(0,0) \rightarrow (\pi,\pi) \rightarrow (\pi,0) \rightarrow (0,0)\}$ in the slab spectrum has $n_F$ odd.
Instead the CTI phase has gapped bulk spectrum and even $n_F$. 
As shown in Fig. \ref{pd2}, switching on this parameter brings about a much richer phase diagram including topological crystalline insulating (CTI) phases for parameters $t^{\prime}_1, t_z^{\prime}$ within the WSM phase for $t_2=0$.  
The nature of this topological insulator is precisely the one discussed in Ref. \onlinecite{fu1} corresponds to a particular case of these phases.  Within these regions, there exist surface states protected by the point  symmetry like the ones shown in Fig. 3. The WSM phase has $n_F=1, \; (n_F=3)$ for $|t^{\prime}_1|> |t^{\prime}_z|$ ($|t^{\prime}_1|< |t^{\prime}_z|$). The thinner TCI phase close to $|t^{\prime}_1|=|t^{\prime}_z|$ has $n_F=2$ while
the other TCI phase has $n_F=4$.

 \section{Summary and discussion}
 
 \label{sum}
 
 We have studied the spectral properties of a tight-binding model for spinless electrons in a double-layer 3D structure with two type of orbitals per unit cell. 
 This model 
 is similar to that
 considered by Fu in Ref. \onlinecite{fu1} as a prototype for a crystalline topological insulator. We have introduced a natural generalization for the interlayer hopping parameters that properly takes into account the symmetry properties of the two orbitals involved. We have shown that 
 in the limit where the intra-layer next-nearest neighbor hopping parameters vanish the low energy spectrum can be effectively described by a generalized 
 Weyl Hamiltonian. That model could be analytically solved and we found that a Weyl semimetal phase takes place for a wide range of parameters. We then showed
 that the topological crystalline insulating phase proposed by Fu, emerges when the intra-layer next-nearest neighbor hopping is switched on. 
 
 In conclusion we have shown that the complete phase diagram of the model is very rich, containing Weyl semimetal as well as normal insulating and crystalline insulating phases. Unlike most of the Weyl semimetal phases analyzed in the literature, the present one does not rely nor on spin orbit interaction neither
 on external magnetic fields. The spin actually does not play any role in driving the different phases of the present 
 model. 
 
 A crucial ingredient leading to the description in terms of the underlying effective Weyl Hamiltonian 
 seems to be a different relative sign in the two intra-layer nearest-neighbor hopping parameters 
 $t_1^A$ and $t_1^B$. 
 To finalize, we would like to comment on the scenario where such change in the sign of the hopping $t_1^a, \;\; a= A, B$ could take place.
 To this end let us first notice that, using the sign convention of Section \ref{Model}, the direct  hopping between $d_{\alpha z}$ (or $p_{\alpha}$) orbitals along
 the $\alpha$ direction is always positive. However, if an intermediate atomic orbital exist (like the circles in the sketch of Fig. 1) the sign could change, depending on the symmetry and the state of charge of the latter. 
 A typical example can be found in  planes of  perovskites, in which case the effective hopping $t_1^a$ between $d_{\alpha z}$ orbitals is originated by the hybridization (named $t_{pd}$) of these orbitals with intermediate 
 $p_{\alpha}$ ones.
 As a consequence, the effective hopping between neighboring $d_{\alpha z}$ is $t_1^a= t_{pd}^2/\Delta$, where $\Delta$ is the charge-transfer energy between the $d_{\alpha z}$ and the intermediate
 $p_{\alpha}$ orbital. Hence, if the intermediate orbitals are empty, the effective hopping $t_1^a$ is negative, while it is positive otherwise. If the intermediate orbitals were $s$, the sign of $t_1^a$ would be, instead,
 negative or positive, depending on whether the intermediate orbital is empty or occupied. Therefore different physical situations can exist in which the 
 effective hopping $t_1^a$ has opposite sign in the two layers.

 \section{Acknowledgements}
 We thank support from CONICET, ANPCyT and UBACyT, Argentina. 
 LA thanks Ruben Weht for many stimulating discussions as well the hospitality of CNEA Constituyentes.

\end{document}